\input harvmac
\input amssym
\noblackbox

\newcount\figno

\figno=0
\def\fig#1#2#3{
\par\begingroup\parindent=0pt\leftskip=1cm\rightskip=1cm\parindent=0pt
\baselineskip=11pt
\global\advance\figno by 1
\midinsert
\epsfxsize=#3
\centerline{\epsfbox{#2}}
\vskip 12pt
\centerline{{\bf Figure \the\figno} #1}\par
\endinsert\endgroup\par}
\def\figlabel#1{\xdef#1{\the\figno}}
\def\pano{\par\noindent}

\def\meno{\medskip\noindent}

\def\fig#1#2#3{
\par\begingroup\parindent=0pt\leftskip=1cm\rightskip=1cm\parindent=0pt
\baselineskip=11pt
\global\advance\figno by 1
\midinsert
\epsfxsize=#3
\centerline{\epsfbox{#2}}
\vskip 12pt
\centerline{{\bf Figure \the\figno} #1}\par
\endinsert\endgroup\par}
\font\cmss=cmss10
\font\cmsss=cmss10 at 7pt

\def\rlx{\relax\leavevmode}
\def\inbar{\vrule height1.5ex width.4pt depth0pt}
\def\IC{\relax\,\hbox{$\inbar\kern-.3em{\rm C}$}}
\def\IR{\relax{\rm I\kern-.18em R}}
\def\IN{\relax{\rm I\kern-.18em N}}
\def\IP{\relax{\rm I\kern-.18em P}}

\def\ZZ{\rlx\leavevmode\ifmmode\mathchoice{\hbox{\cmss Z\kern-.4em Z}}
 {\hbox{\cmss Z\kern-.4em Z}}{\lower.9pt\hbox{\cmsss Z\kern-.36em Z}}
 {\lower1.2pt\hbox{\cmsss Z\kern-.36em Z}}\else{\cmss Z\kern-.4em Z}\fi}

\def\narrowplus{\kern -.04truein + \kern -.03truein}
\def\narrowminus{- \kern -.04truein}
\def\narrowminussub{\kern -.02truein - \kern -.01truein}
\def\cl{\centerline}

\def\ts{\textstyle}

\def\o#1{\overline{#1}}

\def\sqr#1#2{{\vcenter{\vbox{\hrule height.#2pt
 \hbox{\vrule width.#2pt height#1pt \kern#1pt
 \vrule width.#2pt}\hrule height.#2pt}}}}
\def\square
 {\mathop{\mathchoice{\sqr{12}{15}}{\sqr{9}{12}}{\sqr{6.3}{9}}{\sqr{4.5}{9}}}}
\def\spin7{$Spin(7)$}


\def\hepth#1{\hbox{hep-th/#1}}

\lref\rgomis{J. Gomis, {\it D-Branes, Holonomy and M-Theory},
Nucl.Phys. B606 (2001) 3, \hepth{0103115}.}

\lref\rblbr{R. Blumenhagen and V. Braun, {\it 
Superconformal Field Theories for Compact $G_2$ Manifolds},
\hepth{0110232}.}

\lref\rwalcher{ R. Roiban and J. Walcher, {\it
Rational Conformal Field Theories With $G_2$ Holonomy},
\hepth{0110302}.}

\lref\regush{T. Eguchi and Y. Sugawara, {\it
 String Theory on $G_2$ Manifolds Based on Gepner Construction}, 
\hepth{0111012}.}

\lref\rsethi{S. Sethi, C. Vafa and E. Witten, {\it
Constraints on Low-Dimensional String Compactifications},
Nucl.Phys. {\bf B480} (1996) 213, \hepth{9606122}.}

\lref\ratiyah{M. Atiyah and E. Witten, {\it
 M-Theory Dynamics On A Manifold Of $G_2$ Holonomy}, \hepth{0107177}.}

\lref\rwittena{E. Witten, {\it Anomaly Cancellation On Manifolds Of 
 $G_2$ Holonomy}, \hepth{0108165}.}

\lref\rwittenb{E. Witten, {\it Phase Transitions In M-Theory And F-Theory}, 
Nucl. Phys. {\bf B471} (1996) 195, \hepth{9603150}.}

\lref\rbobby{B. Acharya and E. Witten, {\it Chiral Fermions from Manifolds of
    $G_2$ Holonomy}, \hepth{0109152}.}

\lref\ramv{M.~Atiyah, J.~Maldacena, C.~Vafa, 
  {\it An M-theory Flop as a Large N Duality}, 
  \hepth{0011256}.
}

\lref\rvafwit{C. Vafa and E. Witten, {\it A One-Loop Test of String 
Dualities}, Nucl. Phys. {\bf B447} (1995) 261, \hepth{9505053}.}

\lref\rdas{K. Dasgupta and S. Mukhi, {\it A Note on Low-Dimensional String
Compactifications}, Phys.Lett. {\bf B398} (1997) 285, \hepth{9612188}.}

\lref\rbecker{K.~Becker, 
  {\it A Note on Compactifications on \spin7-Holonomy Manifolds}, 
  JHEP {\bf 0105} (2001) 003, 
  \hepth{0011114}.
}

\lref\rvafada{F. Cachazo, K. Intriligator and C. Vafa, {\it 
A Large N Duality via a Geometric Transition}, Nucl.Phys. {\bf B603} (2001) 3,
\hepth{0103067}.
}

\lref\rvafadb{F. Cachazo, S. Katz and C. Vafa, {\it
Geometric Transitions and N=1 Quiver Theories}, 
\hepth{0108120}.
}

\lref\rvafadc{F. Cachazo, B. Fiol, K. Intriligator, S. Katz and C. Vafa,
{\it A Geometric Unification of Dualities}, 
\hepth{0110028}.
}

\lref\raga{M.~Aganagic and C.~Vafa, 
  {\it Mirror Symmetry and a $G_2$ Flop},
  \hepth{0105225}.
}

\lref\ragb{M.~Aganagic and C.~Vafa, 
  {\it $G_2$ Manifolds, Mirror Symmetry and Geometric Engineering},
  \hepth{0110171}.
}

\lref\rmajum{J.~Majumder, 
  {\it Type IIA Orientifold Limit of M-Theory on 
  Compact Joyce 8-Manifold of \spin7-Holonomy}, 
  \hepth{0109076}.
}

\lref\rvafa{C. Vafa, {\it Superstrings and Topological Strings at Large N},
\hepth{0008142}.}

\lref\rtatar{J.D. Edelstein, K. Oh and R. Tatar, {\it 
Orientifold, Geometric Transition and Large N Duality for SO/Sp Gauge
Theories}, JHEP {\bf 0105} (2001) 009, \hepth{0104037}.} 

\lref\rjoyce{D. Joyce, {\it Compact Manifolds of Special Holonomy},
(Oxford University Press, 2000).}

\lref\rpope{M. Cvetic, G.W. Gibbons, H. Lu and C.N. Pope, {\it 
New Complete Non-compact \spin7{} Manifolds}, \hepth{0103155}.}

\lref\rmoore{J.~A.~Harvey and G.~Moore, 
  {\it Superpotentials and Membrane Instantons}, 
  \hepth{9907026}.
}

\lref\rbrandh{A. Brandhuber, J. Gomis, S.S. Gubser and S. Gukov, {\it
Gauge Theory at Large N and New $G_2$ Holonomy Metrics}, 
Nucl.Phys. {\bf B611} (2001) 179,  \hepth{0106034}.}

\lref\rnunez{Jose D. Edelstein and Carlos Nunez, {\it
D6 branes and M theory geometrical transitions from gauged supergravity},
JHEP {\bf 0104} (2001) 028, \hepth{0103167}.}

\lref\rhern{R. Hernandez, {\it Branes Wrapped on Coassociative Cycles},
\hepth{0106055}.}

\lref\rwald{J.P. Gauntlett, N. Kim, D. Martelli and D. Waldram, {\it
Fivebranes Wrapped on SLAG Three-Cycles and Related Geometry},
\hepth{0110034}.}

\lref\rradu{ K. Dasgupta, K. Oh and R. Tatar, {\it
Geometric Transition, Large N Dualities and MQCD Dynamics},
Nucl. Phys. {\bf B610} (2001) 331, \hepth{0105066};
K. Dasgupta, K. Oh and R. Tatar, {\it Open/Closed String Dualities and 
Seiberg Duality from Geometric  Transitions in M-theory}, \hepth{0106040};
K. Dasgupta, K. Oh, J. Park and R. Tatar, {\it 
Geometric Transition versus Cascading Solution},  \hepth{0110050}.}

\lref\racharya{B. S. Acharya, {\it On Realising N=1 Super Yang-Mills in M
    theory}, \hepth{0011089}.}

\lref\rgukov{S. Gukov and J. Sparks, {\it M-Theory on \spin7{} Manifolds: I},
 \hepth{0109025}.} 

\lref\rbwiss{
R.~Blumenhagen and A.~Wi{\ss}kirchen,
  {\it Exactly solvable $(0,2)$ supersymmetric string vacua with GUT
  gauge groups},
  Nucl. Phys. {\bf B454} (1995) 561, \hepth{9506104};
R.~Blumenhagen, R.~Schimmrigk and A.~Wi{\ss}kirchen,
  {\it The $(0,2)$ exactly solvable structure of chiral rings,
  Landau--Ginzburg theories and Calabi--Yau manifolds},
  Nucl.~Phys.~{\bf B461} (1996) 460, \hepth{9510055};
R.~Blumenhagen, R.~Schimmrigk and A.~Wi{\ss}kirchen,
  {\it $(0,2)$ Mirror Symmetry}, Nucl.~Phys.~{\bf B486} (1997) 598, 
  \hepth{9609167}.
}

\lref\rsy{A.~N.~Schellekens and S.~Yankielowicz,
  {\it New modular invariants for $N=2$ tensor products and
  four-dimensional strings},
  Nucl.~Phys.~{\bf B330} (1990) 103.
}

\lref\rwissk{M. Lynker, R. Schimmrigk and A. Wi\ss kirchen, {\it
Landau-Ginzburg Vacua of String, M- and F-Theory at c=12},
Nucl. Phys. {\bf B550} (1999) 123, \hepth{9812195}.}

\lref\rgepner{D.~Gepner,
  {\it Space--time supersymmetry in compactified string theory and
  superconformal models},
  Nucl.~Phys.~{\bf B296} (1988) 757.
}

\lref\rgq{D.~Gepner and Z.~Qiu,
  {\it Modular invariant partition functions for para\-fermionic field
  theories},
  Nucl.~Phys.~{\bf B285} (1987) 423.
}

\lref\rlgcy{D.~Gepner,
  {\it Exactly solvable string compactification on manifolds of $SU(n)$
  holonomy},
  Phys.~Lett.~{\bf B199} (1987) 380.
}

\lref\reot{T.~Eguchi, H.~Ooguri, A.~Taormina and S.~K.~Yang,
  {\it Superconformal algebras and string compactification on manifolds
  with $SU(n)$ holonomy}, 
  Nucl.~Phys.~{\bf B315} (1989) 193.
}

\lref\rschell{
A.~N.~Schellekens and S.~Yankielowicz, 
  {\it Extended chiral algebras and modular invariant 
  partition functions},
  Nucl.~Phys.~{\bf B327} (1989) 673; 
A.~N.~Schellekens and S.~Yankielowicz, 
  {\it Simple currents, modular invariants and Fixed Points},
  Int.~J.~Mod.~Phys.~{\bf A5} (1990) 2903.
}

\lref\rvascha{S.~L.~Shatashvili and C.~Vafa, 
  {\it  Superstrings and Manifolds of Exceptional Holonomy}, 
  \hepth{9407025}.
}

\lref\rjose{J.~M.~Figueroa-O'Farrill, 
  {\it A note on the extended superconformal
  algebras associated with manifolds of exceptional holonomy}, 
  Phys.~Lett.~{\bf B392} (1997) 77, 
  \hepth{9609113}.
}

\lref\rsugi{K.~Sugiyama and  S.~Yamaguchi, 
  {\it Cascade of Special Holonomy
  Manifolds and Heterotic String Theory},  
  \hepth{0108219}.
}

\lref\reguchi{T.~Eguchi and  Y.~Sugawara, 
  {\it CFT Description of String
  Theory Compactified on Non-compact Manifolds with $G_2$ Holonomy}, 
  \hepth{0108091}.
}

\lref\rkachru{S.~Kachru and J.~McGreevy, 
  {\it M-theory on Manifolds of $G_2$
  Holonomy and Type IIA Orientifolds}, 
  JHEP~{\bf 0106} (2001) 027, 
  \hepth{0103223}.
}

\lref\rpioline{H.~Partouche and B.~Pioline, 
  {\it Rolling among $G_2$ vacua}, 
  JHEP~{\bf 0103} (2001) 005, 
  \hepth{0011130}.
}
 
\lref\rkehagiasa{P.~Kaste, A.~Kehagias and H.~Partouche, 
  {\it Phases of supersymmetric gauge theories from M-theory 
  on $G_2$ manifolds},
  JHEP~{\bf 0105} (2001) 058, 
  \hepth{0104124}.
}

\lref\rkehagiasb{A.~Giveon, A.~Kehagias and H.~Partouche, 
  {\it Geometric Transitions, Brane Dynamics and Gauge Theories}, 
  \hepth{0110115}.
}

\lref\rtown{G. Papadopoulos and  P.K. Townsend, {\it 
Compactification of D=11 supergravity on spaces of exceptional holonomy},
Phys. Lett. {\bf B357} (1995) 300, \hepth{9506150}.
}

\lref\rpapa{P.S. Howe and  G. Papadopoulos, 
{\it W Symmetries of a Class of D=2 N=1 Supersymetric 
Sigma Models}, Phys. Lett. {\bf B267} (1991) 362;
P.S. Howe and  G. Papadopoulos, {\it 
Holonomy groups and W-symmetries}, Commun. Math. Phys. {\bf 151} (1993) 467,
\hepth{9202036}. 
}

\Title{\vbox{
 \hbox{HU--EP--01/48}
 \hbox{hep-th/0111048}}}
{\vbox{\centerline{Superconformal Field Theories for Compact} 
\vskip 0.4cm
      {\centerline{Manifolds with \spin7{} Holonomy} }
}}
\centerline{Ralph Blumenhagen \footnote{$^1$}{{\tt e-mail:
 blumenha@physik.hu-berlin.de}} and 
Volker Braun \footnote{$^2$}{{\tt e-mail: volker.braun@physik.hu-berlin.de}}
} 
\bigskip
\centerline{\it Humboldt-Universit\"at zu Berlin, Institut f\"ur  
Physik,}
\centerline{\it Invalidenstrasse 110, 10115 Berlin, Germany}
\smallskip
\bigskip
\centerline{\bf Abstract}
\noindent
We present a  construction of superconformal field theories
for manifolds with \spin7{} holonomy. 
Geometrically these models correspond to the realization of
\spin7{} manifolds as anti-holomorphic quotients of
Calabi-Yau fourfolds. Describing the fourfolds as  Gepner models and
requiring anomaly cancellation we determine  
the resulting Betti numbers
of the \spin7{} superconformal field theory. 
As in the $G_2$ case, we find that the Gepner model and the geometric result
disagree.

\bigskip

\Date{11/2001}
\newsec{Introduction}

Mainly during the last year we have seen an intensified interest into
 the  features of M-theory respectively string theory 
on manifolds of exceptional holonomy 
\refs{\rjoyce\rtown\rmoore\racharya\rbecker\rpioline\rgomis\rpope\rnunez
\rkachru\rkehagiasa\ragb\rbrandh\rhern\ratiyah\rwittena\rgukov\rmajum\rbobby
\rwald\rkehagiasb-\raga}. 
In particular, transitions in the geometry of M-theory on non-compact 
$G_2$ manifolds give rise to non-trivial dualities in 
${\cal N}=1$ supersymmetric gauge theories
\refs{\rvafa\ramv\rvafada\rtatar\rradu\rvafada
\rvafadb-\rvafadc}.

Besides all these successes it is also of importance to study
M-theory and string theory on  compact manifolds
of $G_2$ and \spin7{} holonomy.
Starting with the early work of Shatashvili and
Vafa \rvascha, recently there have been a couple of papers
about the features of the superconformal fields theories
describing strings moving on manifolds of exceptional holonomy
\refs{\rpapa\rjose\reguchi\rsugi\rblbr\rwalcher-\regush}.

In \rblbr\  superconformal fields theories (SCFTs) have been constructed
describing strings on $G_2$ manifolds of the form 
$(CY_3\times S^1)/\ZZ_2$, where
the Calabi-Yau threefold was realized as a Gepner model. 
Discussing three examples in detail it was found that
the $\ZZ_2$ twisted sectors in the SCFT were different from
what one expects in the geometric large volume phase. 
This was explained by the appearance of a discrete NS-NS two-form
flux in the Gepner model\footnote{$^1$}{ 
The partition functions for all Gepner models were constructed very recently
in \regush.}.

In this paper we generalize the construction to the case of manifolds
with \spin7{} holonomy. Starting with the Gepner model for
a Type IIB string on a Calabi-Yau fourfold, we show that, dividing
by an anti-holomorphic $\ZZ_2$ symmetry, one gets a SCFT satisfying all
the requirements for a compactification on a \spin7{} manifolds. 
Requiring  cancellation of the gravitational anomaly in the low
energy two-dimensional theory we will derive  the Betti numbers
of all \spin7{} manifolds, realized as anti-holomorphic quotients
of Gepner models, in terms of the Hodge numbers of the underlying
Calabi-Yau fourfold. Similar to the $G_2$ case, the SCFT results 
disagree in all cases with the supergravity expectation. 
It was suggested that the resolution of this puzzle is that the 
SCFT and the supergravity limit
lie on disconnected
branches in the stringy moduli space.   

This paper  is organized as follows. In section 2 we discuss
Gepner models with $c=12$ describing certain points in the
moduli space of Calabi-Yau fourfolds. In particular, we compute 
from the SCFT the massless spectrum and show that it automatically
satisfies anomaly cancellation. 
In section 3 we divide these Gepner models by an anti-holomorphic
involution, determine the $\ZZ_2$ twisted sector partition function
for the $(2)^2 (6)^4$ Gepner model in detail and compute
the massless spectrum for all possible Gepner models by
employing anomaly cancellation. We also compare the SCFT result
to the geometric spectrum and give arguments for their
discrepancy. Finally,  section 4 contains our conclusions.

\newsec{Gepner models for Calabi-Yau fourfolds}

In this paper we use the same notation as in \refs{\rblbr,\rbwiss}
 and for a general
introduction to Gepner models we refer the reader to the original
literature \refs{\rgepner,\rgq}.

It is well known \refs{\rvafwit,\rsethi}
that purely geometric Type IIA compactifications
on Calabi-Yau fourfolds are destabilized 
by the one loop term
\eqn\desta{  -\int B\wedge X_8(R) ,}
where $B$ denotes the NS-NS two-form and $X_8(R)$ a special quartic polynom 
in the curvature of the fourfold.
For non-vanishing $\chi/24=\int_{CY_4} X_8$ the term \desta\ 
generates a  tadpole for the NS-NS two-form
$B$. Thus, for getting stable vacua one has to either introduce fundamental
strings in the background or turn on some fluxes on the Calabi-Yau fourfolds.
In order to avoid this, we will consider only Type IIB compactifications
in the following. As shown in \rdas, such models are consistent and in 
particular automatically lead to anomaly free models in two space-time
dimensions.  

Some new features appear in two-dimensions: First there exist
Majorana-Weyl spinors which are either right-moving (chiral) or
left-moving (antichiral). Similarly one has to distinguish between
chiral and anti-chiral bosons as well.  Second, in light-cone gauge it
is not immediately clear how one can determine the two-dimensional
massless spectrum directly from the string theoretic partition
function. However, by formally introducing characters of $SO(0)$ we
will argue that it is indeed still possible.

Compactifying the Type IIB string on a Calabi-Yau fourfold  yields
${\cal N}=(0,4)$ supersymmetry in two dimensions. 
The massless spectrum consists of certain numbers of chiral supermultiplets
$(4\phi^+,4\psi^+)$, anti-chiral bosons $\phi^-$ and 
anti-chiral fermions $\psi^-$. 

In the remainder of this section we will discuss how these multiplicities 
can be determined from the Gepner model partition function. 
The Gepner model is constructed from a tensor product
of $N$ unitary models of the ${\cal N}=2$ Virasoro algebra with central charge
$c=12$. In the bosonic string this internal sector is combined
with the $SO(8)\times E_8$ current algebra at level $k=1$. 
The conformal dimensions and the 
$U(1)$ charges of the four highest weight representations of $SO(4n)$ are
 displayed in Table 1.
\vskip 0.8cm
\vbox{
\centerline{\vbox{
\hbox{\vbox{\offinterlineskip
\def\tablespace{height2pt&\omit&&\omit&&
 \omit&\cr}
\def\tablerule{\tablespace\noalign{\hrule}\tablespace}

\hrule\halign{&\vrule#&\strut\hskip0.2cm\hfill #\hfill\hskip0.2cm\cr
& reps.  && $h$ && $q$   &\cr
\tablerule
& $O_{2n}$  && $0$ && $0$  &\cr
\tablespace
& $V_{2n}$  && ${1\over 2}$ && $1$  &\cr
\tablespace
& $S_{2n}$  && ${n\over 4}$ && $0$  &\cr
\tablespace
& $C_{2n}$  && ${n\over 4}$ && $1$  &\cr
}\hrule}}}}
\centerline{
\hbox{{\bf Table 1:}{\it ~~ SO(4n) representations}}}
}
\vskip 0.5cm 
\noindent
Moreover, from the modular $S$-matrix
\eqn\smatrix{
           S={1\over 2}\left(\matrix{  1 & 1 & 1 & 1 \cr
                                       1 & 1 & -1 & -1 \cr
                                       1 & -1 & 1 & -1 \cr
                                       1 & -1 & -1 & 1 \cr}\right) }
one derives the fusion rules shown in Table 2. 
\vskip 0.5cm
\vbox{\cl{\vbox{\offinterlineskip
\def\tablespace{height2pt&\omit&&\omit&&\omit&&\omit &\cr}
\def\tablerule{\tablespace\noalign{\hrule}\tablespace}
\halign{\vrule\quad\hfill#\hfill&\strut\vrule#&\quad#\hfill\quad&\quad 
      #\hfill\quad&\quad#\hfill\quad&\quad#\hfill\quad\vrule\cr
\noalign{\hrule}
$$  && $O$ & $V$ & $S$ & $C$ \cr
\noalign{\hrule}
$O$ && $O$ & $V$ & $S$ & $C$ \cr
$V$ && $V$ & $O$ & $C$ & $S$ \cr
$S$ && $S$ & $C$ & $O$ & $V$ \cr
$C$ && $C$ & $S$ & $V$ & $O$ \cr
\noalign{\hrule}
}}}
\centerline{
\hbox{{\bf Table 2:}{\it ~~ SO(4n) fusion rules}}}
}
\meno
Note in particular, that the fusion of the spinor respectively antispinor
representation with itself gives the singlet representation.  
The Gepner model is constructed by starting with the diagonal partition
function and then taking orbits under both 
the simple current \refs{\rschell,\rsy} implementing the GSO projection
\eqn\simple{ 
        J_{GSO}=(0,1,1)^N\otimes (S_8) }
and the $N$ simple currents
\eqn\simpleb{ 
        J_{a}=\prod_{i=1}^{a-1} (0,0,0)\otimes (0,0,2)
               \otimes \prod_{i=a+1}^{N} (0,0,0)  \otimes (V_8)  }
ensuring  that from
the individual factor theories only NS respectively R  sector states are 
combined. After formally applying the bosonic string map exchanging
 $SO(8)\times E_8$ and
$SO(0)$ representations in the following way
\eqn\bosonic{(O_8,V_8,S_8,C_8)\to (V_0,O_0,-C_0,-S_0) }
one gets the partition function of the $c=12$ Gepner model.

A Calabi-Yau fourfold has four non-trivial Hodge numbers
$h_{31}$, $h_{21}$, $h_{11}$ and $h_{22}$ satisfying the relation
\eqn\hodge{ h_{22}=44+4\, h_{11} +4\, h_{31} -2\, h_{21}. }   
Can one detect these numbers in the chiral ring of the Gepner model?
In Table 3 we summarize how the massless  orbits in the Gepner model
are connected to the Hodge numbers of the Calabi-Yau manifold.
\vskip 0.8cm
\vbox{
\centerline{\vbox{
\hbox{\vbox{\offinterlineskip
\def\tablespace{height2pt&\omit&&\omit&&
 \omit&\cr}
\def\tablerule{\tablespace\noalign{\hrule}\tablespace}

\hrule\halign{&\vrule#&\strut\hskip0.2cm\hfill #\hfill\hskip0.2cm\cr
& Hodge  && left && right   &\cr
\tablerule
& $h_{31}$  && $\left({1\over 2},\mp 1\right)_{NS}\otimes O_0-
                \left({1\over 2},\pm 1\right)_R\otimes S_0$ && 
               $\left({1\over 2},\mp 1\right)_{NS}\otimes O_0-
                \left({1\over 2},\pm 1\right)_R\otimes S_0$  &\cr
\tablespace
& $h_{11}$  && $\left({1\over 2},\mp 1\right)_{NS}\otimes O_0-
                \left({1\over 2},\pm 1\right)_R\otimes S_0$ && 
               $\left({1\over 2},\pm 1\right)_{NS}\otimes O_0-
                \left({1\over 2},\mp 1\right)_R\otimes S_0$  &\cr
\tablespace
& $h_{21}$  && $\left({1\over 2},\mp 1\right)_{NS}\otimes O_0-
                \left({1\over 2},\pm 1\right)_R\otimes S_0$ && 
               $\left({1},\mp 2\right)_{NS}\otimes V_0-
                \left({1\over 2}, 0\right)_R\otimes C_0$  &\cr
\tablespace
& $h_{22}$  && $\left({1},\mp 2\right)_{NS}\otimes V_0-
                \left({1\over 2}, 0\right)_R\otimes C_0$ && 
               $\left({1},\mp 2\right)_{NS}\otimes V_0-
                \left({1\over 2}, 0\right)_R\otimes C_0$  &\cr
}\hrule}}}}
\centerline{
\hbox{{\bf Table 3:}{\it ~~ massless orbits}}}
}
\vskip 0.5cm 
Note, that the equation \hodge\ is automatically satisfied
in the Gepner model. 
Since the character of the vector representation of $SO(0)$ is 
formally vanishing (a vector in two dimensions has no on-shell
degrees of freedom), there do not appear any massless NS states
in the orbits related to $h_{21}$ and $h_{22}$. 
Beside the matter multiplets there also appear R-R massless states 
in the vacuum orbit
\eqn\vacuum{\eqalign{ &\left[\left({0},0\right)_{NS}\otimes V_0-
                \left({\ts{1\over 2}},2\right)_R\otimes C_0 -
              \left({\ts{1\over 2}}, -2\right)_R\otimes C_0\right]_L\times\cr
              &\left[\left({0}, 0\right)_{NS}\otimes V_0-
                \left({\ts{1\over 2}}, 2\right)_R\otimes C_0 -
                \left({\ts{1\over 2}}, -2\right)_R\otimes C_0\right]_R .\cr}}
The Type IIB massless spectrum can now be computed by applying the 
rules shown in Table 4.
\vskip 0.8cm
\vbox{
\centerline{\vbox{
\hbox{\vbox{\offinterlineskip
\def\tablespace{height2pt&\omit&&\omit&&
 \omit&\cr}
\def\tablerule{\tablespace\noalign{\hrule}\tablespace}

\hrule\halign{&\vrule#&\strut\hskip0.2cm\hfill #\hfill\hskip0.2cm\cr
& sector   && left-right SO(0) comb. && massless state    &\cr
\tablerule
& NS-NS  && $(O_0)_L\times (O_0)_R$ && 1 non-chiral boson  &\cr
\tablerule
& R-R  && $(S_0)_L\times (S_0)_R$ && 1 (anti)-chiral boson &\cr
\tablespace
& R-R  && $(C_0)_L\times (C_0)_R$ && 1 (anti)-chiral boson  &\cr
\tablerule
& R-R  && $(S_0)_{L,R}\times (C_0)_{R,L}$ &&  $-$ &\cr
\tablerule
& NS-R && $(O_0)_{L,R}\times (S_0)_{R,L}$  && 1 chiral fermion & \cr
\tablerule
& NS-R && $(O_0)_{L,R}\times (C_0)_{R,L}$  && 1 anti-chiral fermion & \cr
\tablerule
& NS-R && $(V_0)_{L,R}\times (S_0,C_0)_{R,L}$  && $-$ & \cr
}\hrule}}}}
\centerline{
\hbox{{\bf Table 4:}{\it ~~ chiral massless states}}}
}
\vskip 0.5cm 
Neglecting the dilaton which is part of the supergravity multiplet
and taking the four (anti-)chiral bosons from the vacuum orbit
into account the total number of (anti-)chiral bosons is
\eqn\bostot{  n_{\phi^+}+n_{\phi^-}=6\, h_{31}+6\, h_{11}\, + h_{22} +4.}
For the number of chiral respectively antichiral fermions one  
obtains
\eqn\fermtot{\eqalign{  n_{\psi^+}&=4\left( h_{31}+h_{11}\right) \cr
                       n_{\psi^-}&=4\, h_{21} . \cr }}
Since 4 chiral fermions and 4 chiral bosons form one ${\cal N}=(0,4)$
supermultiplet, one can deduce the number of chiral and antichiral
bosons separately
\eqn\bostotb{\eqalign{  n_{\phi^+}&=4\left( h_{31}+h_{11}\right) \cr
                       n_{\phi^-}&= 6\, h_{31}+6\, h_{11}-
                  2\, h_{21}+48, \cr }}
where we have used \hodge.
The gravitational anomaly for ${\cal N}=(0,4)$ supergravity theory
in two dimensions is proportional to
\eqn\anomal{ 4+{n_{\phi^+}-n_{\phi^-} \over 12 }+ 
          {n_{\psi^+}-n_{\psi^-} \over 24}, }
where the first term comes from the gravitinos.
Indeed, for the massless spectrum shown in \fermtot\ and
\bostotb\ the anomaly vanishes as it should be for 
a modular invariant stringy partition function. 

\newsec{Gepner models for \spin7{} manifolds}

Given a Gepner model with $c=12$, we now divide by the anti-holomorphic
involution $\sigma^*$ which geometrically acts as complex conjugation. 
Therefore, the supersymmetry on the world-sheet is broken
to ${\cal N}=1$. 
In complete analogy to \rblbr, in the Gepner model $\sigma^*$ acts
as charge conjugation in each individual ${\cal N}=2$ tensor factor.
Thus, in the ${\scriptstyle\sigma^*}\square\limits_1$ sector of the orbifold
partition function only uncharged states can contribute.
As opposed to the $G_2$ case, for all levels $k_i$ even
there do not only exist uncharged states in the NS-NS sector but
also in the R-R sector. Let us discuss this case first.

\subsec{All levels even}

Using the action of the anti-holomorphic involution on the
${\cal N}=2$ unitary models with level $k=1,2,3,6$ as derived
in \rblbr, we have computed the orbifold of the $(2)^2 (6)^4$ Gepner model
in detail. Geometrically, the $(2)^2 (6)^4$ Gepner model
corresponds to the Calabi-Yau fourfold $\IP_{1,1,1,1,2,2}[8]$ with
Hodge numbers $h_{31}=443$, $h_{11}=1$, $h_{21}=0$ and $h_{22}=1820$.

The computation for the partition function
is absolutely analogous to the $G_2$ examples discussed in \rblbr.
We merely  use this concrete example
 as a guiding model for the
general results presented in the following.

The uncharged states in the NS-NS sector, which are contained in
the highest weight representations (HWR) $(l,m,s)=(l,0,0)$ of the
${\cal N}=2$ unitary models contribute to the 
${\scriptstyle\sigma^*}\square\limits_1$ sector of the $\ZZ_2$ orbifold
in the same way as in \rblbr.
However,  as the $(2)^2 (6)^4$ example 
\eqn\insera{\eqalign{
 { {\scriptstyle\sigma^*}\square\limits_1 }^{\, NS}
   ={1\over 2} \left| V_0 \right|^2\, \sum_{i,j=0}^1 \sum_{k,l,m,n=0}^3 
\Biggl| &\psi^{2i}(\sigma^*)\,\psi^{2j}(\sigma^*)\biggl(
            \chi^{2k}(\sigma^*)\, \chi^{2l}(\sigma^*)\, \chi^{2m}(\sigma^*)\, 
            \chi^{2n}(\sigma^*)+ \cr
         &\chi^{6-2k}(\sigma^*)\, \chi^{6-2l}(\sigma^*)\, 
            \chi^{6-2m}(\sigma^*)\, 
            \chi^{6-2n}(\sigma^*)  \biggr) \Biggr|^2 \cr }}
reveals in detail the NS-NS sector alone
does only yield half-integer coefficients in the 
twisted sector partition function, which would not allow
the interpretation as a trace over states in a Hilbert space.
In \insera\ we have used the notation from \rblbr.

Since the state  $\prod_i({k_i\over 2},{k_i \over 2}+1,1)\otimes (C_0)$ 
in the Ramond sector is uncharged, the R-R sector also contributes
to  ${\scriptstyle\sigma^*}\square\limits_1$.
Taking this into account, for the  $(2)^2 (6)^4$  Gepner model one gets
\eqn\inserb{\eqalign{
 {{\scriptstyle\sigma^*}\square\limits_1}^{\,R}
   = 2\, \left| C_0 \right|^2\,
\left\vert \, \left(\psi^{1}(\sigma^*)\right)^2
      \, \left(\chi^{3}(\sigma^*)\right)^4 \right|^2, }}
where the extra factor of two signals that the relevant orbit
has half the length of the standard orbit. 
The $k=2$ and $k=6$ characters in \inserb\ are
\eqn\chart{\eqalign{
        \psi^{1}(\sigma^*)&=\sqrt{2\eta\over \theta_2}\, 
                  \sqrt{\theta_2 \over 2\eta} \cr
         \chi^{3}(\sigma^*)&=\sqrt{2\eta\over \theta_2}\, {1\over 2}\,
      \left( \chi^R_{3\over 32} -\chi^R_{65\over 32} \right), \cr}}
using the fact that the ${\cal N}=2$ unitary model at level
$k=6$ 
can be expressed as the unitary model of the ${\cal N}=1$ super Virasoro
algebra at level $m=6$.

A modular $S$-transformation
to the $\sigma^*$ twisted sector yields 
\eqn\parti{\eqalign{
  1\square\limits_{\sigma^*}=\left\vert \sqrt{\eta\over\theta_4}\right\vert^6
{1\over 2}\biggl[I_1(q,\o q)\, \vert O_0+V_0-S_0-C_0 \vert^2 +
                 I_2(q,\o q)\, \vert O_0-V_0-S_0+C_0 \vert^2\biggr] ,}}
where the first term is the modular transform of the NS-NS sector
contribution to ${\scriptstyle\sigma^*}\square\limits_1$ and the second
term the R-R sector contribution.
The series $I_1$ and $I_2$ start like                 
\eqn\twisted{\eqalign{ 
    I_1(q,\o q)&=\left| \psi_0^2\, \left(\chi_{1\over 32}^{NS}\right)^4 
            \right| +\ldots \cr
    I_2(q,\o q)&=\left| \psi_0^2\, 
\left(\chi_{1\over 32}^{\widetilde{NS}}\right)^4    \right| +\ldots .\cr}}
The complete twisted sector \parti\
has indeed integer coefficient and therefore allows the interpretation
as a trace. Moreover, the twisted sector satisfies level matching and is
free of tachyons. It turns out that the ground state in 
the twisted sector is massless.

However, before determining the massless spectrum in the twisted
sector let us compute  the massless spectrum from the untwisted
sector not only for the $(2)^2 (6)^4$ Gepner model but quite
generally for every Gepner model with even levels. 
First, since $\sigma^*$ acts as charge conjugation, the
vacuum orbit tells us that supersymmetry is broken to
${\cal N}=(0,2)$. 

Moreover, neglecting the dilaton we get 
two (anti-)chiral bosons from the  vacuum orbit. 
The first  and second class of orbits in Table 3 contribute $3h_{31}+3h_{11}$
(anti-)chiral bosons, while as in the fourfold case the third class of
orbits does not lead to any boson at all. 
For the states in the fourth class of orbits, one has to take into 
account that (only) the left-right combination
\eqn\invst{ \left[\prod_i({\ts{k_i\over 2}},{\ts{k_i \over 2}}+1,1)
                \otimes (C_0)\right]_L \times 
            \left[\prod_i({\ts{k_i\over 2}},{\ts{k_i \over 2}}+1,1)
                \otimes (C_0)\right]_R }
is invariant under $\sigma^*$. As mentioned above, this  orbit has 
half the length of the standard orbit and therefore we find that 
the total number of (anti-)chiral bosons arising in the fourth
class of orbits is ${h_{22}\over 2}+1$. 
The overall number of (anti-)chiral bosons is therefore
\eqn\bosganz{  n_{\phi^+}+n_{\phi^-}=3\, h_{31}+3\, h_{11}\, + 
                  {h_{22}\over 2} +3.}
Since precisely half of the fermions are projected out 
we get
\eqn\fermganz{\eqalign{  n_{\psi^+}&=2\left( h_{31}+h_{11}\right) \cr
                       n_{\psi^-}&=2\, h_{21}, \cr }}
so that we 
can deduce the number of chiral and antichiral
bosons separately
\eqn\bosganzb{\eqalign{  n_{\phi^+}&=2\left( h_{31}+h_{11}\right) \cr
                       n_{\phi^-}&= 3\, h_{31}+3\, h_{11}-
                  h_{21}+25. \cr }}
Note that for this spectrum the gravitational anomaly
\eqn\anomalb{ I=2+{n_{\phi^+}-n_{\phi^-} \over 12 }+ 
          {n_{\psi^+}-n_{\psi^-} \over 24} }
does not vanish but gives $I=-{1\over 12}$. 
This should be canceled by the contribution from the massless modes in 
the $\ZZ_2$ twisted sector. 
The partition function \parti\ derived for the $(2)^2 (6)^4$ Gepner 
model gives rise to the massless modes
\eqn\massltw{\eqalign{ n_{\phi^+}^{tw}=2, \quad n_{\phi^-}^{tw}=2 \cr
                       n_{\psi^+}^{tw}=2, \quad n_{\psi^-}^{tw}=0 \cr}}
which precisely contribute the missing states to cancel
the gravitational anomaly. 

Even though we have explicitly computed the twisted sector
partition function only for the $(2)^2\, (6)^4$ Gepner model, 
turning the argument around, 
anomaly cancellation tells us that for all Gepner models
with only even levels there must be the same contribution \massltw\ 
from the $\ZZ_2$ twisted sector\footnote{$^1$}{While this paper
was in preparation we received \regush, which showed quite generally
that for the $G_2$ Gepner
models with only even levels one always finds 2 chiral states from 
the $\ZZ_2$ twisted sector. Apparently, something very similar happens in the
\spin7{} case.}.

By dimensional reduction of Type IIB string theory 
on a \spin7{} manifold we expect 
the number of chiral respectively anti-chiral two dimensional
bosons and fermions to be
\eqn\chiral{\eqalign{  n^{geo}_{\phi^+}&=2\, b_2 + 2\, b^-_4 +2 \cr
                       n^{geo}_{\phi^-}&=2\, b_2 + b_4 +2 \cr
                       n^{geo}_{\psi^-}&=2\, b_3, \cr}} 
where $b_n$ denote the Betti numbers of the \spin7{} manifold:
There are $b_4^-+1$ geometric moduli giving rise to the same
number of non-chiral bosons. The NS-NS and R-R two-forms
yield $2 b_2$ non-chiral bosons. The R-R zero-form gives another
non-chiral boson and the self-dual four-form gives rise to
$b^-_4$ chiral bosons and $b^+_4$ anti-chiral bosons.

Now we can try to compute the Betti numbers of the \spin7{} SCFT model 
in terms of the Hodge numbers of the Calabi-Yau fourfold
we started with.
However, from the SCFT data one can not uniquely determine 
the number of $\sigma^*$ invariant 2-cycles, so that we can only say
that $b_2=h_{11}+1-l$ with $0\le l\le h_{11}+1$. Here
we have also taken the possible contribution from the $\ZZ_2$ twisted
sector \massltw\ into account.
Using $b_3=h_{21}$ we get for the number of self-dual and
anti self-dual 4-cycles
\eqn\fourcyc{\eqalign{  b_2&=h_{11}-l+1 \cr
                        b_3&=h_{21} \cr
                        b_4^-&=h_{31}+l-1 \cr
                       b_4^+&=2\, h_{31} + h_{11} -h_{21} +l +24.}}
The resulting Euler number is
\eqn\eulera{ \chi=2+2\, b_2-2\, b_3 +b_4=3(h_{31}-h_{21}+h_{11}+9) }
which is in particular independent of $l$.
The existence of one covariantly constant spinor on a \spin7{} manifold
implies via the index theorem a relation between the Betti numbers
\eqn\bettirel{  24=b_3-b_2-2\, b_4^- + b_4^+ -1 .}
We consider it as a nice check of   our computation that the Betti numbers
derived from the SCFT analysis indeed automatically satisfy
this condition independently of $l$. 
Thus, up to the parameter $l$ the SCFT determines the Betti numbers
completely, which was though expected from the analysis
in \rvascha. Since the SCFT only fixes  the sums $b_2+b_4^-$ and
$b_2+b_4^+$, it was conjectured in \rvascha\ that this reflects
a generalized mirror symmetry for \spin7{} manifolds. 
All geometric target spaces with Betti numbers as in \fourcyc, for
each $l$ lead to the same SCFT and can therefore not be distinguished by
strings moving in these backgrounds.

\subsec{The large radius geometry}

The geometric phase corresponding to a  Gepner model
with only even levels is a  Calabi-Yau hypersurface 
of the form
\eqn\hyper{  \sum_{i=1}^6 z_i^{k_i+2} =0 }
in an appropriate weighted projective space. 
Thus the $\ZZ_2$ acts freely on the fourfold since
there are no real solutions. Furthermore the involution is
antiholomorphic and  therefore inverts the K\"ahler class
arising from the ambient space. So there
are $h_{11}-l$ invariant $2$--cycles with $0< l \leq h_{11}$
allowing us to determine the remaining Betti numbers. Obviously
$b_0=1$, $b_1=0$, $b_2=h_{11}-l$. Then the involution exchanges
$H^{2,1}\leftrightarrow H^{1,2}$ and therefore $b_3=h_{21}$.

It is not obvious how the involution acts on $H^{2,2}$, but we can
determine this from the Euler number:
\eqn\eulerCY{ 
  \eqalign{
    2+2\, b_2-2\, b_3 +b_4
    &= \chi(Y/\ZZ_2) = {1 \over 2} \chi(Y)
    ={{2+2\, h_{11}-4\, h_{21}+2+2\, h_{31}+h_{22}} \over 2 }
  \cr 
    & \Rightarrow
    b_4 = 2 l -h_{11} + h_{31} + {h_{22} \over 2}
   = 22 + 2 l + h_{11} +3\, h_{31} -\, h_{21}
  }
}
so there are ${h_{22} \over 2} +2l -h_{11}-1$ classes invariant in
$H^{2,2}(Y)$. 
Finally we determine the refined Betti numbers $b_4^\pm$ from
$\hat{A}(Y/\ZZ_2)=1$, that is from eq.~\bettirel. We find
\eqn\refinedbetti{ 
  \eqalign{
    b_4^-&=h_{31}+l-1 \cr
    b_4^+&=2\, h_{31} + h_{11} -h_{21} +l +23.
} }

The quotient $Y/\ZZ_2$ of course has $\pi_1=\ZZ_2$ and therefore not
the full \spin7{} holonomy, but rather 
${\rm Hol}(Y/\ZZ_2)=\ZZ_2\ltimes SU(4)$. However there is one constant
spinor and this suffices for our purposes.

\subsec{SCFT vs. Geometry}

Similar to the $G_2$ Gepner models the SCFT result
disagrees with the large radius computation: From the 
analysis of the geometric phase one would not expect any massless
states in the twisted sector, and indeed the computation of the Betti
numbers misses these degrees of freedom.
We expect the resolution to this puzzle
to be similar to the $G_2$ Gepner models. 
There we argued that the moduli coming from the dimensional reduction
of the NS-NS 2-form are frozen in the $G_2$ manifold, so that
one can not continuously go from the Gepner model to the
geometric phase.

Here if a two-cycle is anti-invariant under the anti-holomorphic 
involution, which is always the case for the
K\"ahler class inherited from the ambient space, 
the $B$ field can still take the 
two discrete values $B=0$ and $B=1/2$. Since generally the Gepner model
is expected to correspond to a point in the moduli where
$B=1/2$, the SCFT model and the supergravity limit lie
on disconnected branches of the moduli space.

\subsec{At least one level odd}

If at least one level of the $c=12$ Gepner model is odd, 
there are no uncharged states in the R-R sector.
Therefore, the massless spectrum in the 
untwisted sector is precisely half the spectrum of the fourfold
Gepner model. 
The number of (anti-)chiral fermions and bosons turns out to be
\eqn\spectot{\eqalign{  n_{\psi^+}&=2\left( h_{31}+h_{11}\right) \cr
                       n_{\psi^-}&=2\, h_{21}  \cr 
                n_{\phi^+}&=2\left( h_{31}+h_{11}\right) \cr
               n_{\phi^-}&= 3\, h_{31}+3\, h_{11}-
                  h_{21}+24, \cr }}
which by themselves  cancel the gravitational anomaly. Therefore, we do not
expect any further massless states in the twisted sector which 
is in accord with the analogous result for the $G_2$ Gepner models 
\refs{\rblbr,\rwalcher,\regush}.
Comparing the SCFT spectrum \spectot\ with the expected
result from dimensional reduction \chiral, one can again deduce 
the Betti numbers of the \spin7{} SCFT
\eqn\fourcycb{\eqalign{ b_2&=h_{11}-l \cr
                        b_3&=h_{21} \cr
                        b_4^-&=h_{31}+l-1 \cr
                       b_4^+&=2\, h_{31} + h_{11} -h_{21} +l +23.}}
Note, that these Betti-numbers are slightly different from the 
ones in \fourcyc, but agree with the geometric expressions \refinedbetti\
derived for a free orbifold action. 
The resulting Euler number is
\eqn\euler{ \chi=2+2\, b_2-2\, b_3 +b_4=3(h_{31}-h_{21}+h_{11}+8) .}

In this case, in the expression for the hypersurface
\eqn\hyper{  \sum_{i=1}^6 z_i^{k_i+2} =0 }
at least one exponent is odd and one would expect non-trivial
fixed loci under complex conjugation. These would lead to
further massless states in the twisted sector, so that also here
the SCFT result and the geometric result disagree.

\newsec{Conclusions}

In this paper  we have generalized the construction of exactly
solvable SCFTs to the case of \spin7{} manifolds. 
Starting with a Gepner model with $c=12$ describing certain 
points in the moduli space of string compactifications
on Calabi-Yau fourfolds, we divided by charge conjugation
to get a \spin7{} manifold. The massless Type IIB spectra
we got satisfy all the conditions we expect for
\spin7{} compactifications. Employing the cancellation
of the gravitational anomaly in two dimensions, it was possible
to compute for all Gepner models the resulting Betti numbers
of the \spin7{} manifold in terms of the Hodge numbers of
the Calabi-Yau fourfold. One might speculate that these
formulae are not only restricted to Gepner models but
might be true even for more general hypersurfaces in weighted
projective spaces respectively for more general orbifolds of 
c=12 Landau-Ginzburg models \rwissk. 

Similar to the $G_2$ case the twisted sectors for the SCFT and
the supergravity computations turned out to be different, reflecting
the fact that both models lie on separate branches of the 
\spin7{} moduli space.

\centerline{{\bf Acknowledgments}}\pano
The group is supported in part by the EEC contract ERBFMRXCT96-0045.

\listrefs

\bye
\end